\documentclass[preprint]{revtex4}
\usepackage{graphicx}
\usepackage{amsmath}

\begin{document}

\title{Revealing the collision energy dependence of $\eta/s$ in RHIC-BES Au+Au collisions using Bayesian statistics}

\author{Jussi Auvinen}
\affiliation{Department of Physics, Duke University, Durham, NC 27708, USA}
\author{Iurii Karpenko}
\affiliation{INFN - Sezione di Firenze, I-50019 Sesto Fiorentino (Firenze), Italy}
\author{Jonah E. Bernhard}
\affiliation{Department of Physics, Duke University, Durham, NC 27708, USA}
\author{Steffen A. Bass}
\affiliation{Department of Physics, Duke University, Durham, NC 27708, USA}

\begin{abstract}
We investigate the collision energy dependence of $\eta/s$
in a transport + viscous hydrodynamics hybrid model.
A Bayesian analysis is performed on RHIC beam energy scan data
for $Au+Au$ collisions at $\sqrt{s_{NN}}=19.6,39,$ and $62.4$ GeV.
The resulting posterior probability distributions
for the model parameters show a preference
for a larger value of $\eta/s$ at $19.6$ GeV compared to $62.4$ GeV,
indicating dependence on baryon chemical potential $\mu_B$.
\end{abstract}

\maketitle

\section{Introduction}
\label{sec:introduction}

Significant progress has been made in the past few years in determining QGP properties,
such as the temperature dependence of shear viscosity over entropy density ratio $\eta/s$
\cite{Csernai:2006zz,Niemi:2015qia,Denicol:2015nhu,Plumari:2016lul}.
However, in a recent RHIC beam energy scan study \cite{Karpenko:2015xea} it was found that
the best fit with the experimental data was reached using larger values of $\eta/s$ at lower
collision energies, suggesting an additional dependency on the baryon chemical potential $\mu_B$.

It is generally difficult to determine the uncertainties associated with the extracted best-fit values of QGP properties,
as the computational models used in the analysis typically have numerous interconnected parameters,
which need to be tuned on large sets of experimental data.
We tackle this issue in the present study by utilizing Bayesian statistics combined with Markov chain Monte Carlo (MCMC)
to calibrate the computational model to data.
The end result of such an analysis is a multidimensional probability distribution,
which provides not only a set of data-calibrated parameter values,
but a full uncertainty quantification as well.
This approach has already been applied with great success to $Pb+Pb$ collisions at the LHC \cite{Bernhard:2016tnd}.

In this article, we investigate the $\mu_B$ dependence of $\eta/s$, using the collision energy $\sqrt{s_{NN}}$ as the control parameter.
We simulate $Au+Au$ collisions at $\sqrt{s_{NN}}=19.6,39,$ and $62.4$ GeV,
using the same transport + (3+1)-D viscous hydrodynamics model as in Ref.~\cite{Karpenko:2015xea}.
In addition of introducing the robust uncertainty quantification method described above,
this revised analysis benefits also from the recently published experimental data
on identified particle multiplicities and mean transverse momentum \cite{Adamczyk:2017iwn}.

\section{Transport + hydrodynamics hybrid model}
\label{sec:model}

In the utilized hybrid model, the initial non-equilibrium evolution is handled by the UrQMD hadron-string cascade \cite{Bass:1998ca,Bleicher:1999xi}.
The transition from hadron-string transport to hydrodynamics happens some time after the two colliding nuclei have passed through each other.
Beyond this condition, the exact transition time is a free parameter of the model.
At the transport-to-hydro transition, the microscopic particle properties are mapped to hydrodynamic densities
using 3-D Gaussians with width parameters $R_{\textrm{trans}}$, $R_{\textrm{long}}$.

The hydrodynamic evolution is done using a (3+1)-D relativistic viscous hydrodynamics code \cite{Karpenko:2013wva}
with a constant value of $\eta/s$, which is given as an input.
A chiral model equation of state \cite{Steinheimer:2010ib} is used during the hydro evolution
to accommodate for finite values of baryon chemical potential.

The transition from hydrodynamics back to hadron transport happens
when the local rest frame energy density in hydro cells falls below the user-defined switching value $\epsilon_{SW}$.
The ''Cornelius'' routine \cite{Huovinen:2012is} is used to construct the iso-energy density hypersurface,
from which hadrons are sampled according to Cooper-Frye procedure.
The rescatterings and resonance decays of the sampled hadrons are processed within UrQMD.

\section{Statistical analysis}
\label{sec:stats}

According to Bayes' theorem, the posterior probability for the model parameters
is a product of our prior knowledge about the plausible range of values for each parameter,
and the likelihood of any given combination of parameter values being the ''correct'' one
when compared to experimental data:
\begin{equation}
\mathcal{L}(x) \propto \exp\left(-\frac{1}{2}(y(x)-y^{\textrm{\,exp}})^T\Sigma^{-1}(x)(y(x)-y^{\textrm{\,exp}}) \right)
\end{equation}
where $\Sigma$ is the covariance matrix,
representing the uncertainties related to the comparison of model output $y(x)$ and the data $y^{\textrm{\,exp}}$
for the input parameter combination $x$.

In practice, we produce samples of the Bayesian posterior distribution using Markov chain Monte Carlo,
where the initial positions of the random walkers are based on the prior
(in this case, a uniform distribution in a 5-dimensional hypercube),
and the likelihood function determines the probability for a walker to accept a proposed step direction in the parameter space.
We use $\mathcal{O}(1000)$ random walkers to produce a sufficient number of samples of the posterior distribution.

The MCMC method necessitates fast evaluations of the likelihood function,
but running the actual hybrid model simulations to determine $y(x)$ for arbitrary $x$
requires an infeasible amount of computational effort.
We circumvent this problem by utilizing Gaussian process (GP) emulators,
which provide an efficient method for predicting $y(x)$ with quantitative uncertainty.

In order to use GP emulators, they need to be conditioned on training data.
For this analysis, approximately 100 training data points were produced for each collision energy.
A Latin hypercube method was used to sample the training points,
to ensure a representative sample of the whole input parameter space.

\section{Results}
\label{sec:results}

The analysis was performed using data for
charged particle multiplicity $N_{ch}$ \cite{Abelev:2008ab}, pseudorapidity distribution $dN_{ch}/d\eta$ \cite{Alver:2010ck},
and elliptic flow $v_2\{\textrm{EP}\}$ ($\sqrt{s_{NN}}=62.4$ GeV) \cite{Adamczyk:2012ku,Magdy:2017kji}
or $v_2\{2\}$ ($\sqrt{s_{NN}}=19.6$ and $39$ GeV) \cite{Adamczyk:2012ku}.
Identified particle observables include charged pion HBT radii $R_{\textrm{out}}$, $R_{\textrm{side}}$ and $R_{\textrm{long}}$ \cite{Adamczyk:2014mxp},
$K^+/\pi^+$ ratio \cite{Abelev:2008ab},
and multiplicities and mean transverse momentum of $\pi^+$, $\pi^-$, $K^+$, $K^-$ \cite{Adamczyk:2017iwn}
and $\Omega$ \cite{Aggarwal:2010ig,Chatterjee:2015fua,Adamczyk:2015lvo}.
Proton data was not included in this analysis, as proton yields suffer from additional uncertainties related to feed-down corrections.

\begin{figure}
 \centering
 \includegraphics[width=4.9cm]{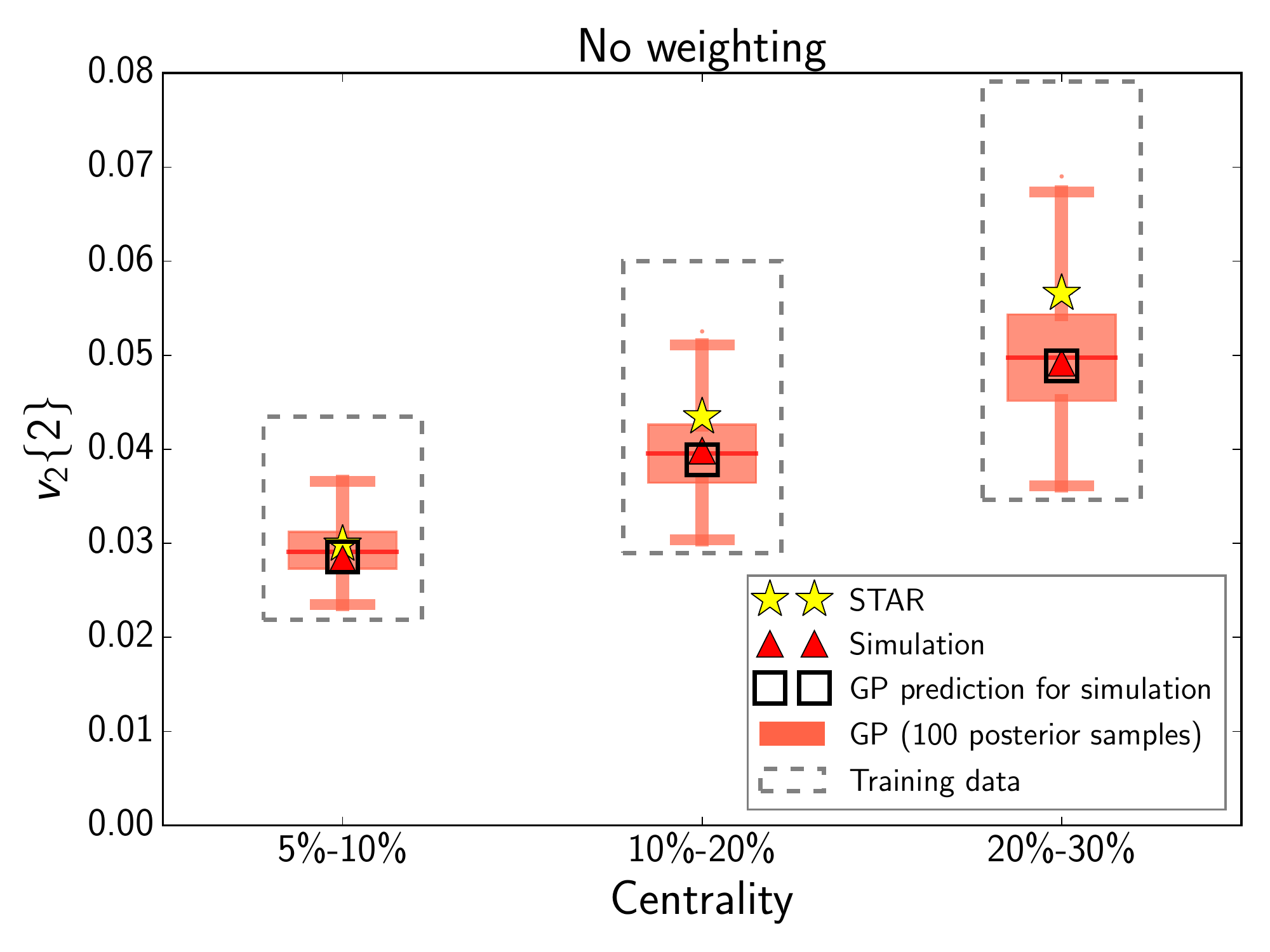}
 \includegraphics[width=4.9cm]{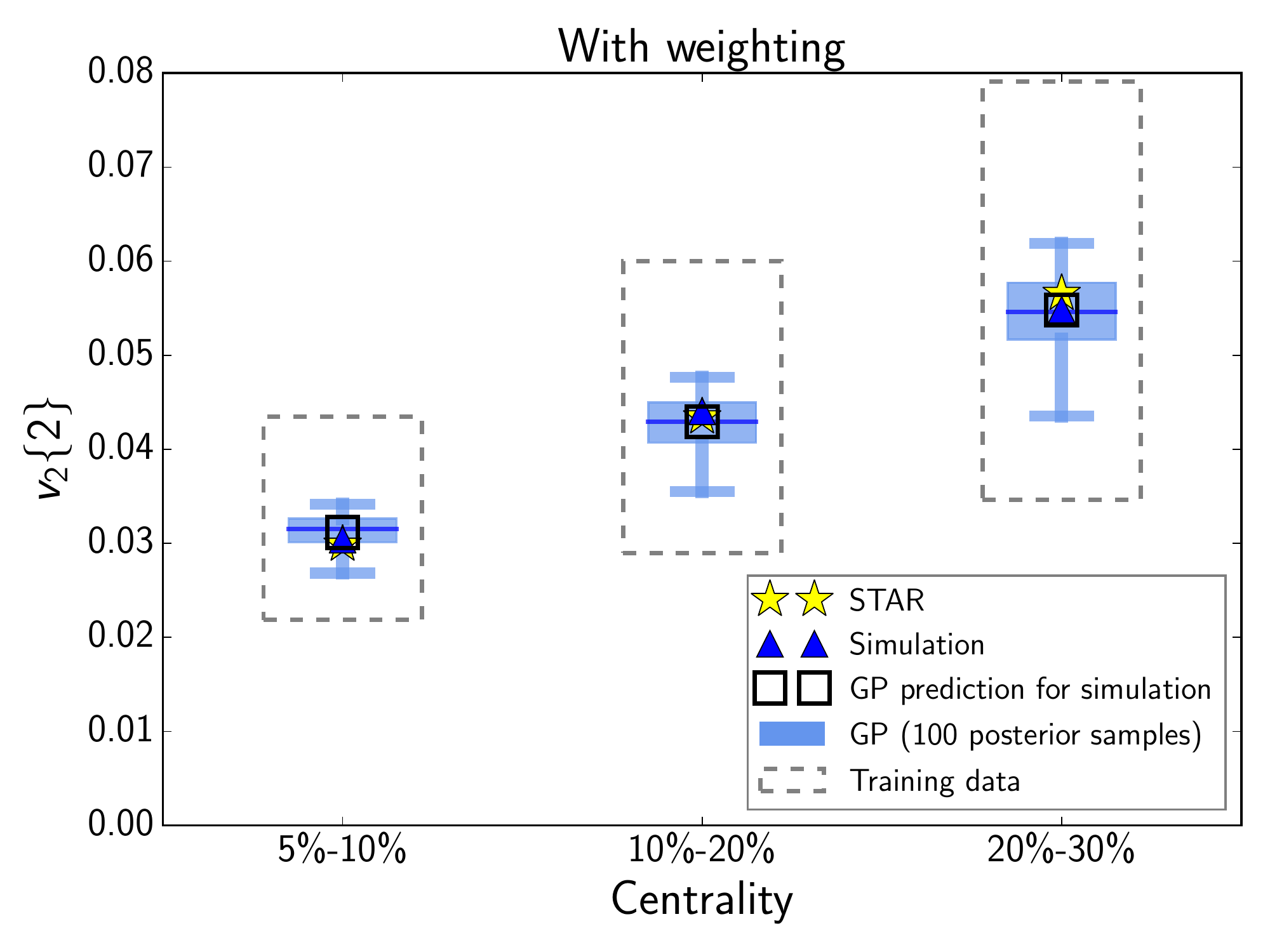}
 \includegraphics[width=4.9cm]{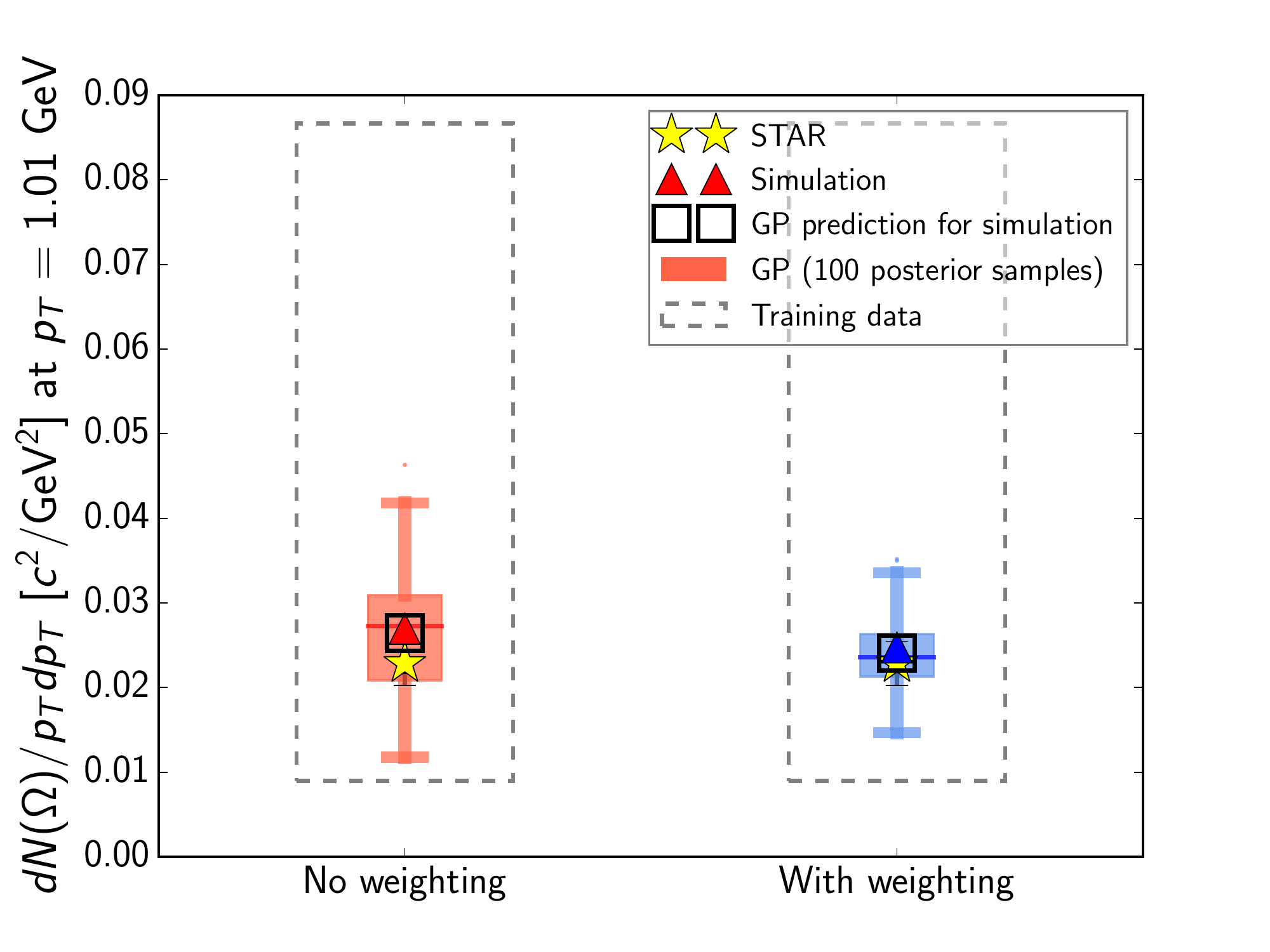}
 \caption{(Color online) Comparison of emulator predictions on elliptic flow $v_2$ and $\Omega$ multiplicity at $\sqrt{s_{NN}}=39$ GeV,
 both without weighting (red) and with additional weights (blue).
 Triangles represent full model calculations using median values of parameter posterior distributions,
 while open squares are the respective emulator predictions for the same input parameter combinations.
 STAR $v_2$ data from \cite{Adamczyk:2012ku} and $\Omega$ data from \cite{Adamczyk:2015lvo}.}
 \label{fig:weights39}
\end{figure}

We have increased the weight of $v_2$ in the analysis by a factor of 5
and $N(\Omega)$ by a factor of 4,
as these observables have been found to improve constraints on $\eta/s$ and $\epsilon_{SW}$,
respectively. As an example, figure \ref{fig:weights39} shows the effect of weighting on $v_2$ and $N(\Omega)$ at $\sqrt{s_{NN}}=39$ GeV.
The range of emulator predictions for model outputs,
based on 100 random draws from the posterior distributions,
is clearly in better agreement with the measured values when extra weighting is introduced.
The emulator quality is demonstrated in the figure by showing both the GP predictions and the real model output,
when the median values of posterior probability distributions were used as input.

\begin{figure}
 \centering
 \includegraphics[width=4.9cm]{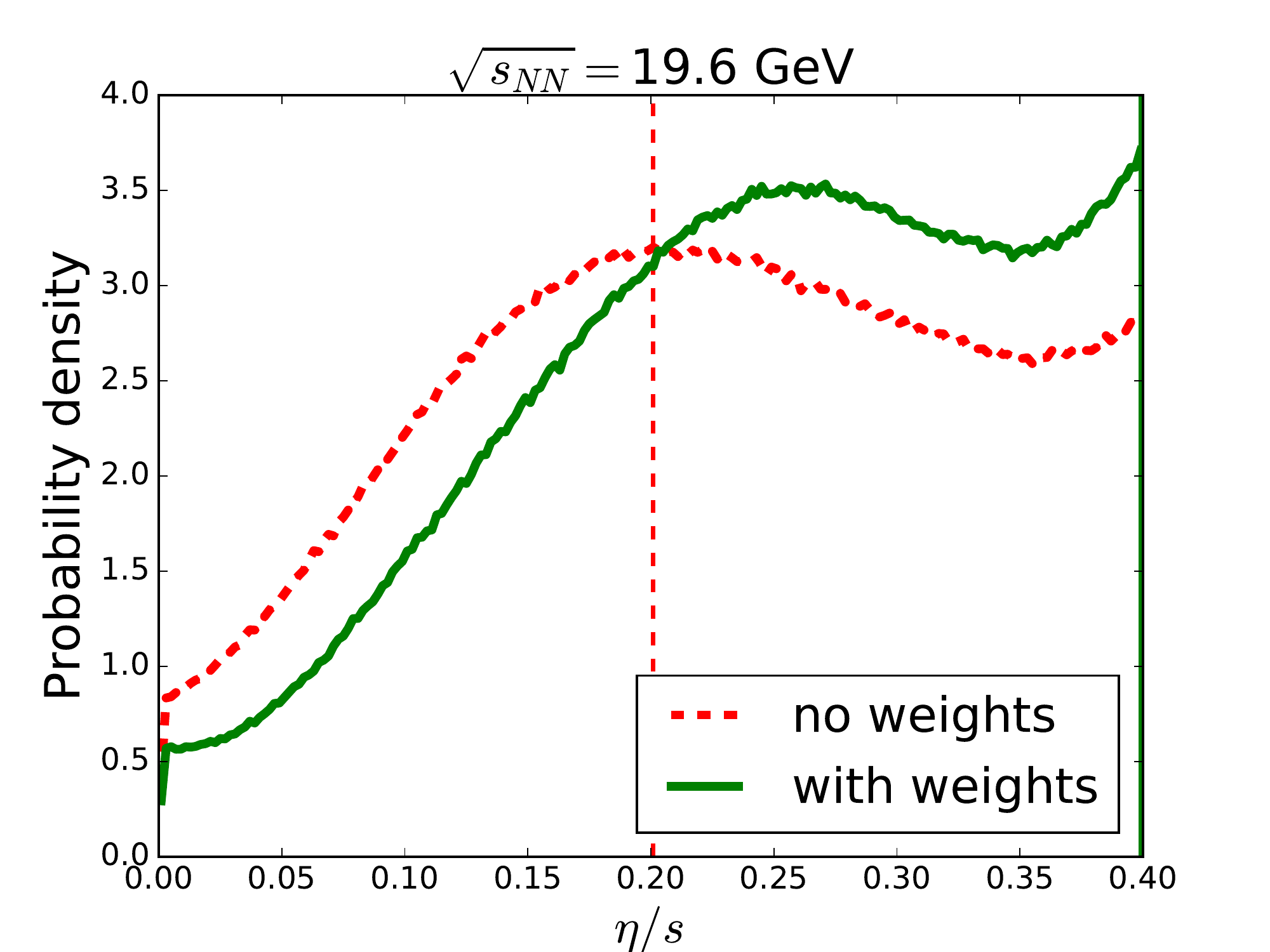}
 \includegraphics[width=4.9cm]{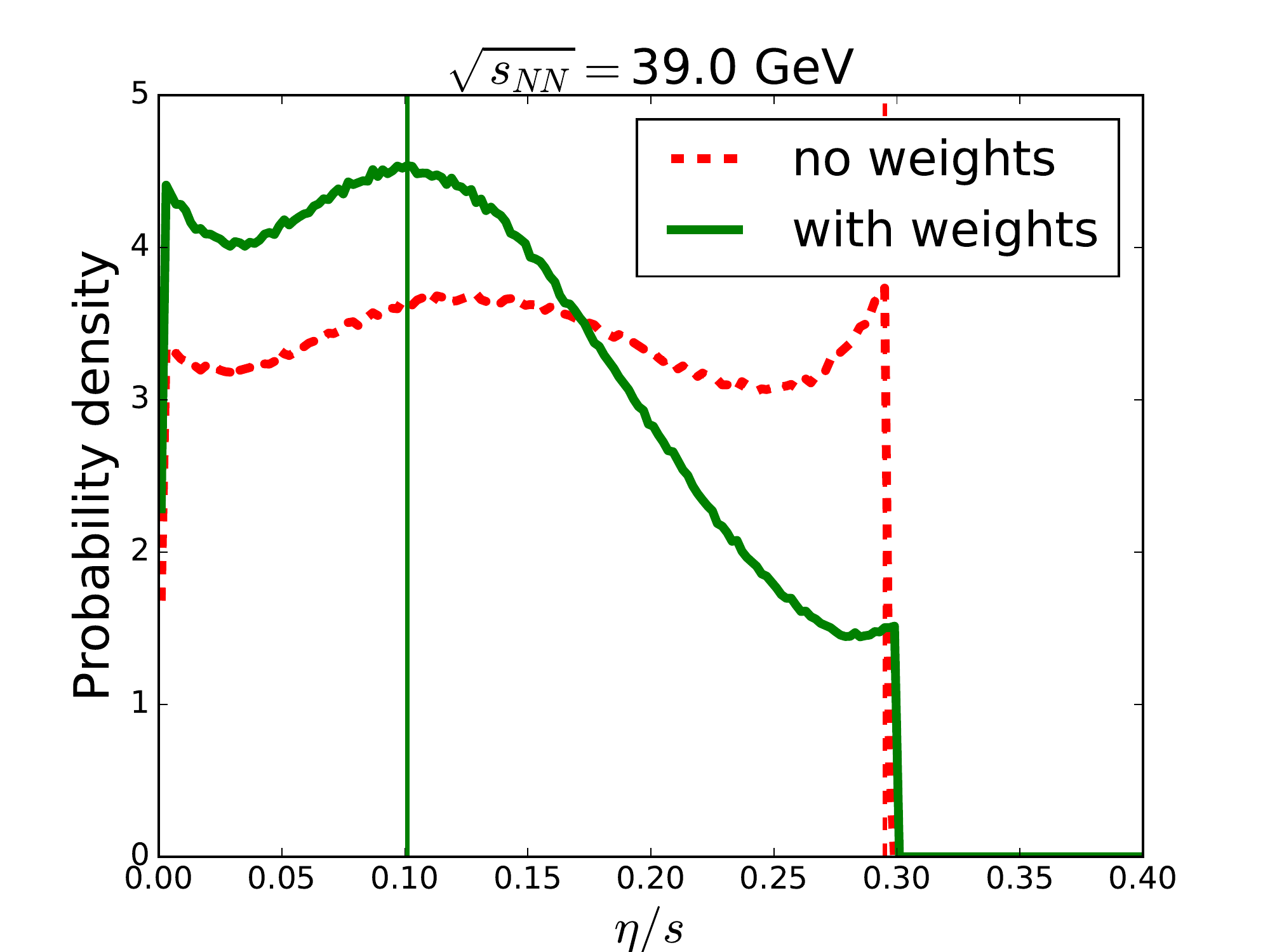}
 \includegraphics[width=4.9cm]{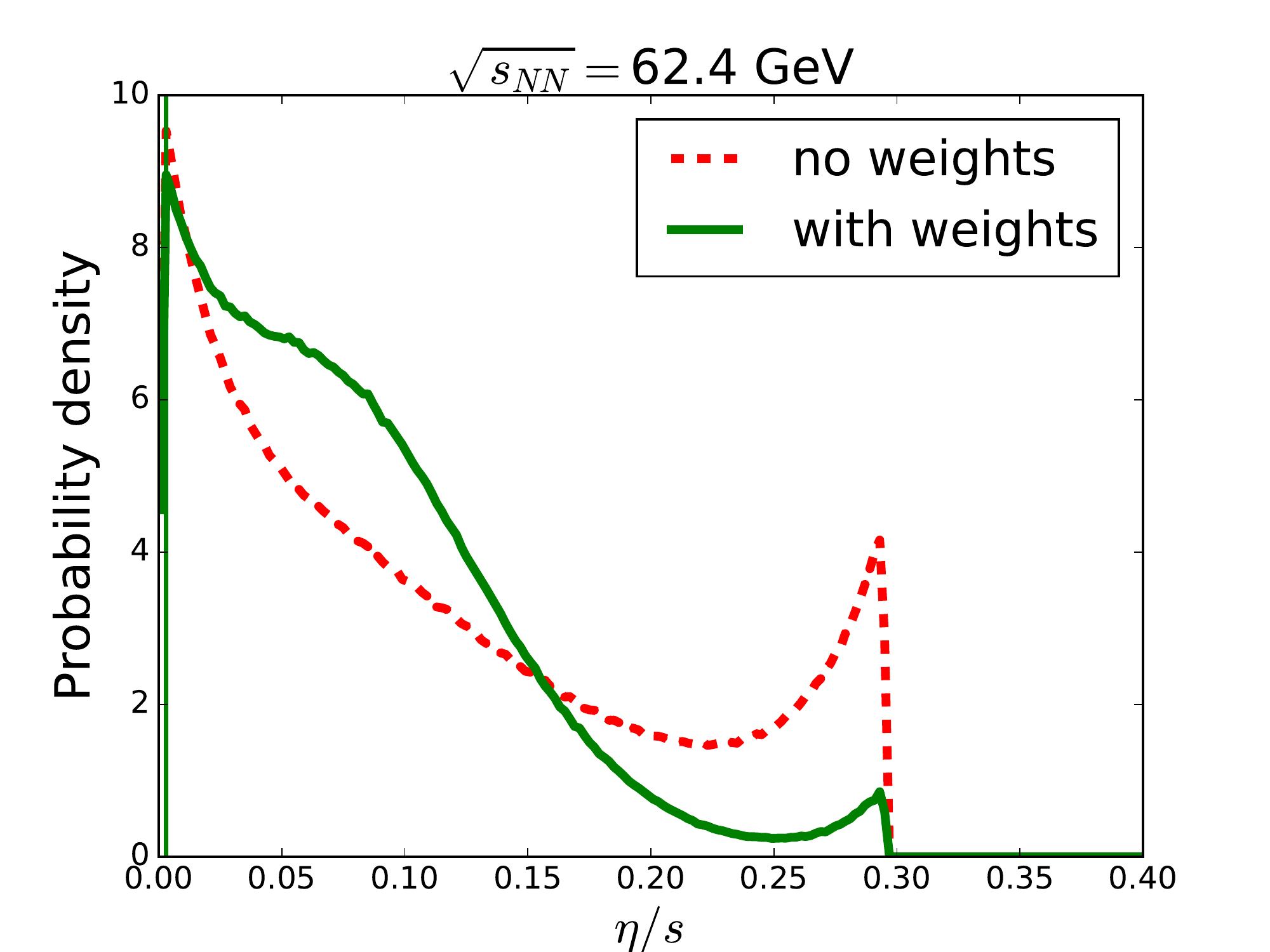}
 \caption{Comparison of shear viscosity over entropy density ratio $\eta/s$ posterior distributions with weighting (solid green curves) and without weighting (dashed red curves).
 Vertical lines indicate the peak values of the distributions.}
 \label{fig:etas_posterior}
\end{figure}

Figure \ref{fig:etas_posterior} shows one-dimensional projections of posterior probability distributions for $\eta/s$ for the three collision energies,
with and without weighting.
If no weighting is introduced, shear viscosity remains largely unrestricted, especially at $\sqrt{s_{NN}}=39$ GeV.
At $\sqrt{s_{NN}}=19.6$ GeV, there is a visible preference towards values over 0.2,
while for $\sqrt{s_{NN}}=62.4$ GeV the probability distribution peaks at $\eta/s \approx 0$.
These tendencies are more emphasized when the importance of $v_2$ and $\Omega$ observables are increased in the analysis,
while the posterior distribution at $\sqrt{s_{NN}}=39$ GeV develops a peak at intermediate value $\eta/s \approx 0.1$.

\begin{figure}
 \centering
 \includegraphics[width=4.9cm]{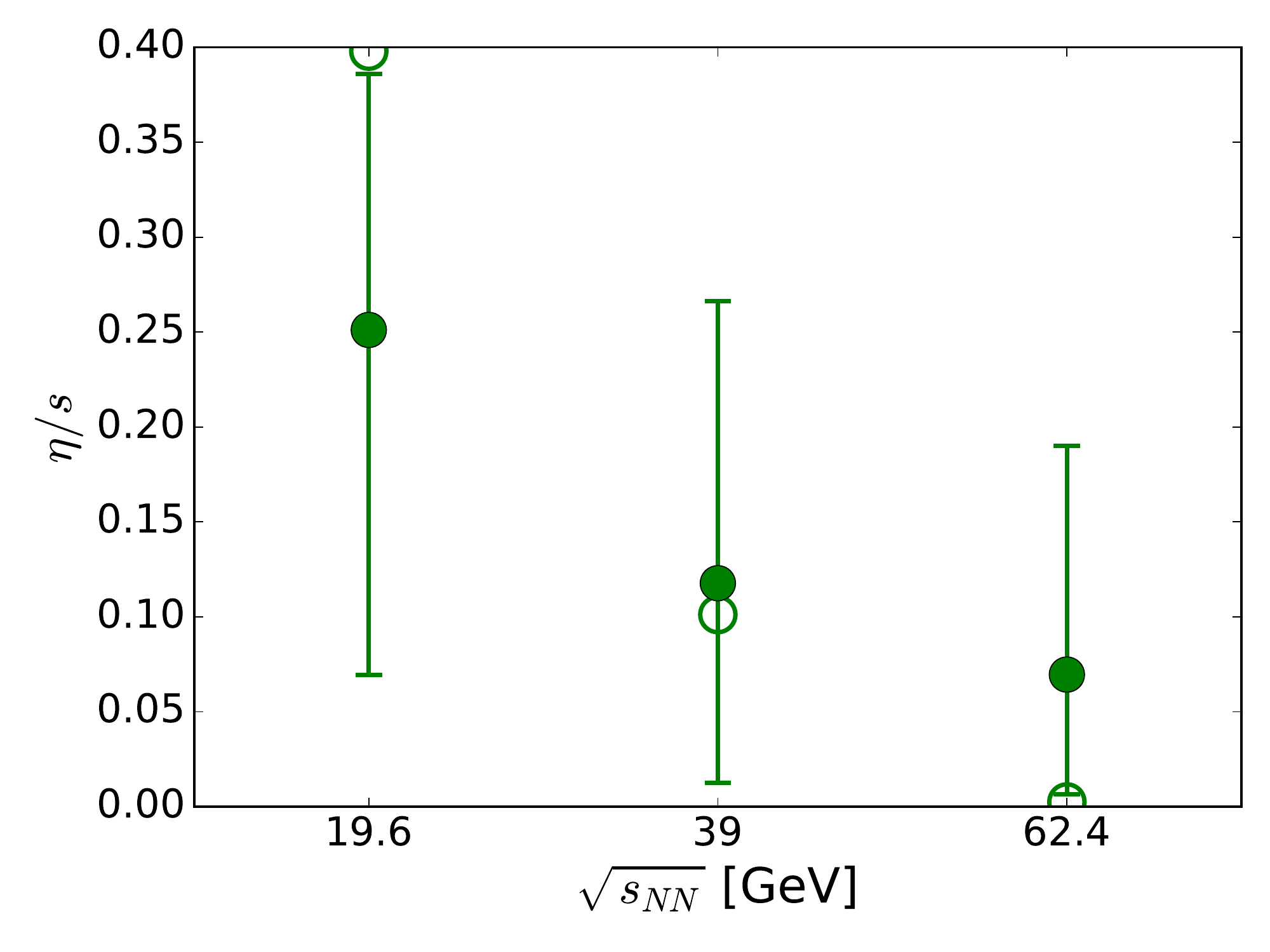}
 \includegraphics[width=4.9cm]{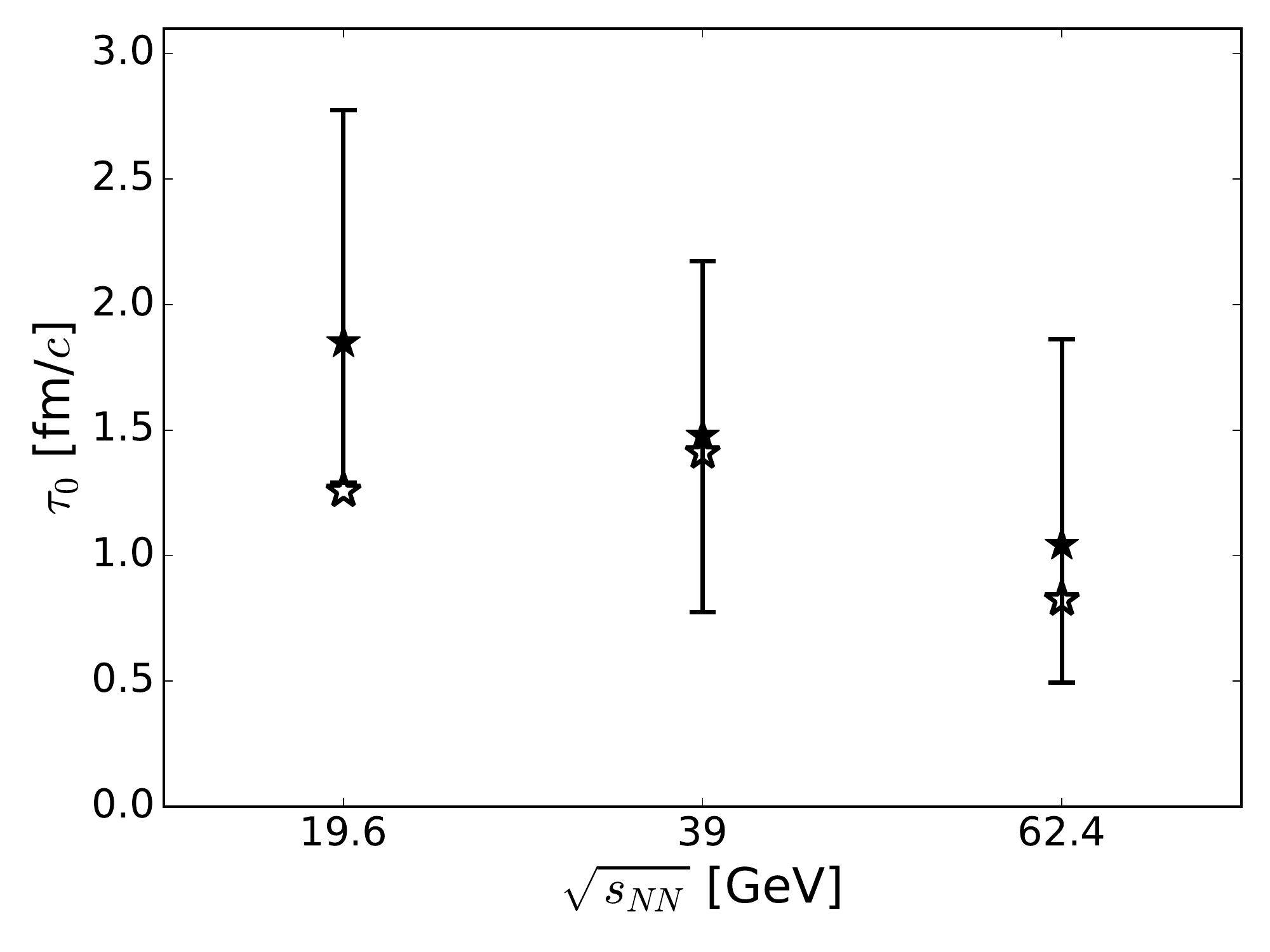}
 \includegraphics[width=4.9cm]{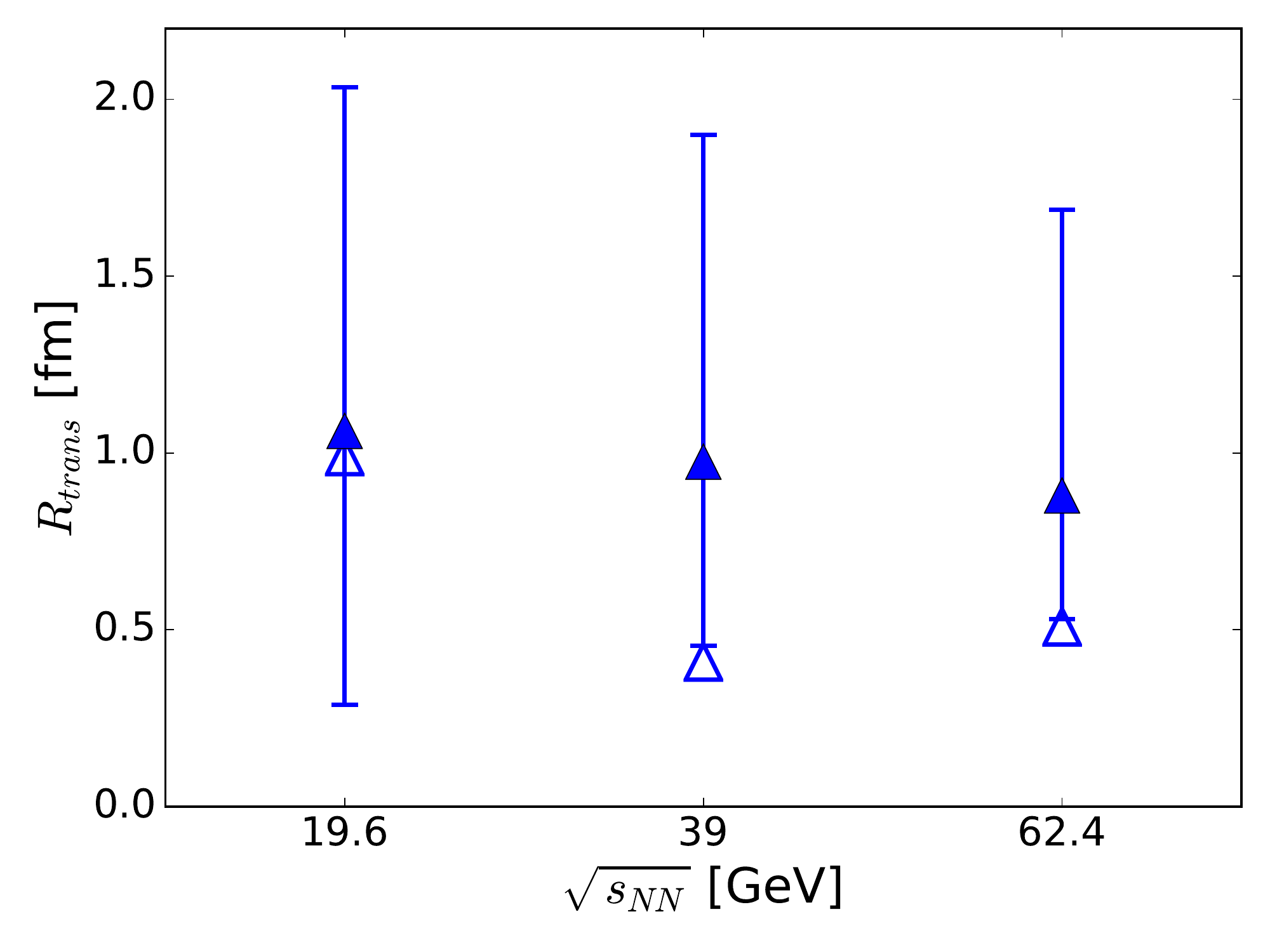}
 \includegraphics[width=4.9cm]{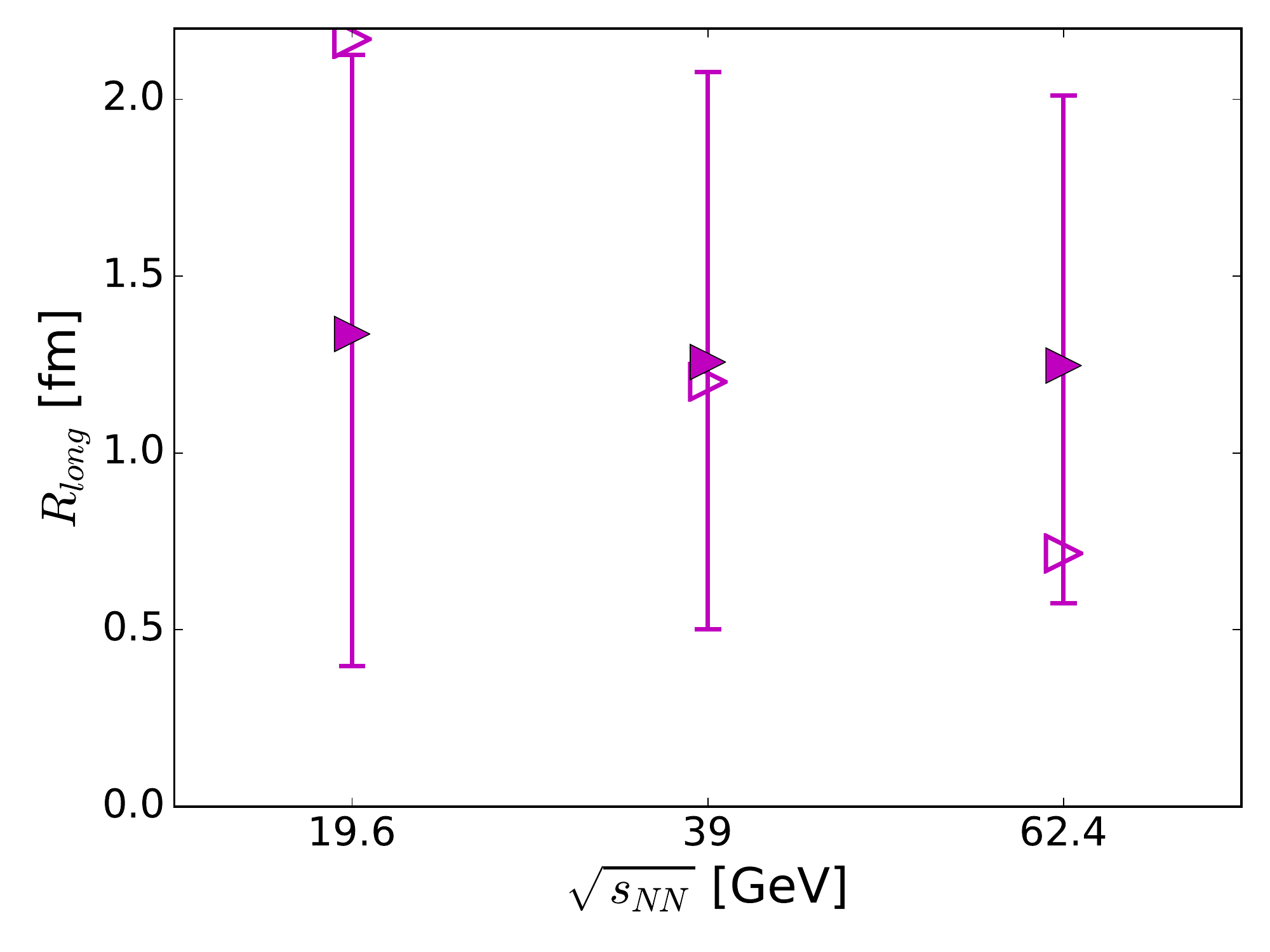}
 \includegraphics[width=4.9cm]{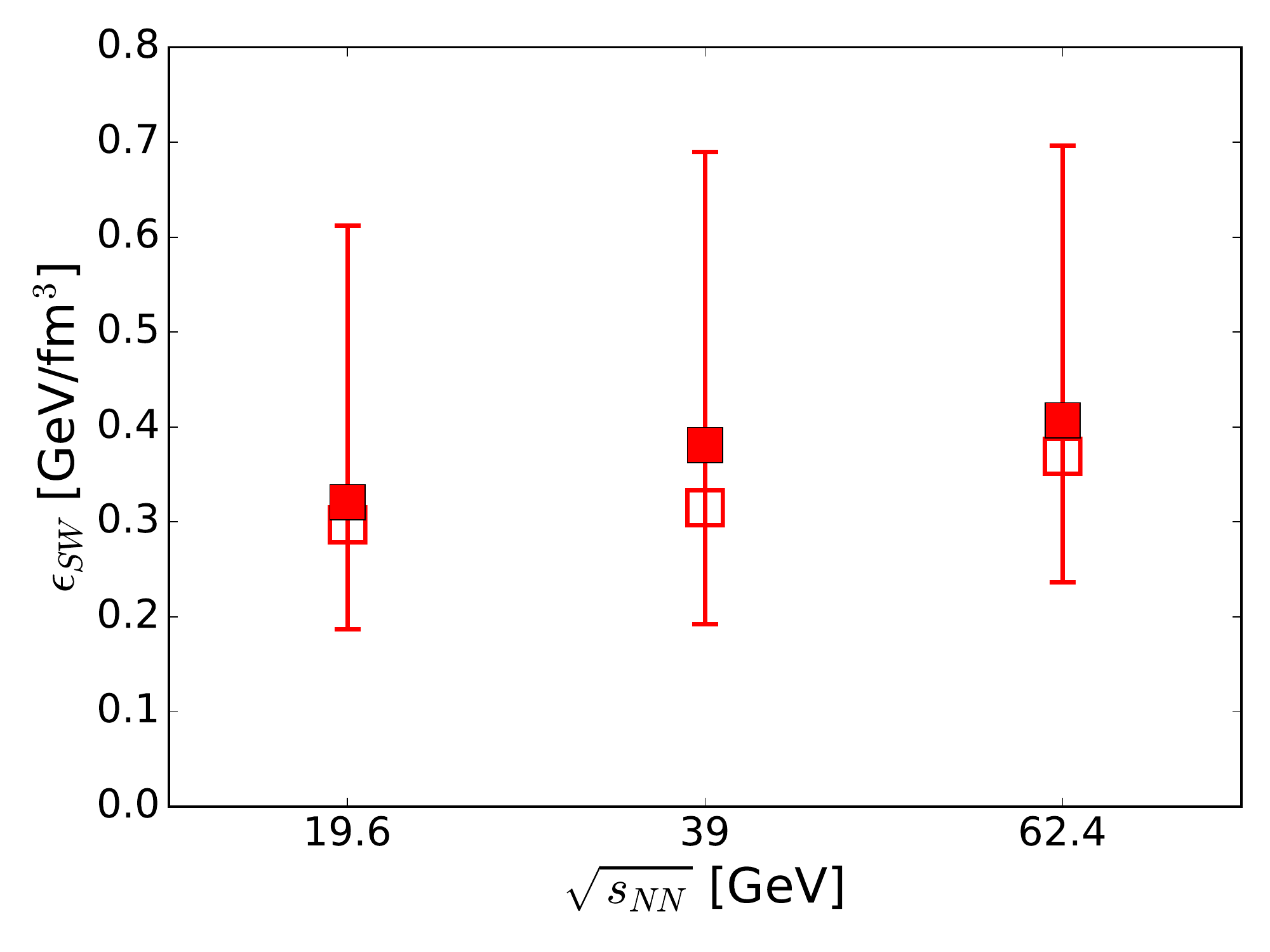}
 \caption{Illustration of the collision energy dependence of 1-D posterior probability distributions for the model parameters in the weighted scenario.
 Open symbols indicate the peak values of the distributions,
 while the full symbols and associated error bars represent median values and 90\% confidence limits, respectively.}
 \label{fig:bes_posteriors}
\end{figure}

We present a rough illustration of the collision energy dependencies of the posterior distributions of all five model parameters in figure \ref{fig:bes_posteriors}.
Even with the introduction of additional weighting, the 90\% confidence limits remain large for all parameters.
Especially the width parameters, controlling the creation of the initial density profile for the hydrodynamics, are poorly constrained by the data.
This uncertainty about the initial state is naturally reflected also on the other parameters.

The differences between peak and median values (open and solid symbols in Fig.~\ref{fig:bes_posteriors}) indicate that the distributions are skewed,
as already illustrated for $\eta/s$ in figure \ref{fig:etas_posterior}.
The consistent right-skewness of transverse smearing $R_{\textrm{trans}}$ suggests that the optimal value is more likely to be 1.0 fm or less.
The strongest statement can be made on the value of hydro-to-transport switching energy density,
which has both the peak and the median on the range $\approx 0.3 - 0.4$ GeV/fm$^3$ for all investigated collision energies,
making it very likely that the optimal value will be found within this range.

\section{Summary}
\label{sec:summary}

Utilizing Bayesian statistics and Gaussian process emulators,
we have performed a state-of-the-art model-to-data comparison
on a transport + hydrodynamics hybrid model describing $Au+Au$ collisions at RHIC beam energy scan.
While the collision energy dependence of the posterior probability distribution strongly suggests
that shear viscosity over entropy density ratio $\eta/s$ depends on baryon chemical potential $\mu_B$,
the present uncertainties prohibit precise statements about the ''correct'' parameter values.
We expect the situation to improve once the dynamics of the initial fluidization are better understood.

\section{Acknowledgements}
\label{sec:acknowledgements}
This work has been supported by NSF grant no.~NSF-ACI-1550225 and by DOE grant no.~DE-FG02-05ER41367.
CPU time was provided by the Open Science Grid, supported by DOE and NSF.

\end{document}